\documentstyle[12pt]{article}
\title{
Canonical Quantization of the Liouville Theory,
Quantum Group Structures, and Correlation Functions\thanks{Talk
presented at the 16th Johns Hopkins Workshop on Current Problems
in Particle Theory, \quad \quad June 8-10, 1992 Gothenburg, Sweden}}
\author{
Gerhard Weigt\\
DESY - Institute for High Energy Physics\\
D-1615 Zeuthen, Platanenallee 6, Germany}
\date{     }

\begin{document}

\pagestyle{plain}
\maketitle

\vspace{0.5cm}

\begin{abstract}
We describe a self-consistent canonical quantization of Liouville theory
in terms of canonical free fields. In order to keep the non-linear
Liouville dynamics, we use the solution of the Liouville equation as a
canonical transformation.
 This  also defines a Liouville vertex operator.
We show, in particular, that a canonical quantized
 conformal and local quantum Liouville theory has a quantum group
structure, and we discuss
correlation functions for non-critical
strings.
\end{abstract}

\vspace{1.5cm}

\noindent
{\Large \bf 1.\ \ Introduction}


\vspace{0.5cm}

Liouville theory arises in mathematical and physical
situations which
involve surfaces. Although its classical dynamics was solved in terms of free
fields a long
time ago [1], there are still many open problems especially in
the quantum case. Recent results of
this theory [2], which follow from the matrix model approach or by
performing path integrals, remain obscure from the point of view of an
exactly solved quantum Liouville theory.

However, it is tempting to assume that the quantum structure of the Liouville
theory is also completely given by free fields, as in the
classical case [3-5]. Then
canonical quantization might be instrumental in understanding the quantum
Liouville dynamics. To simplify the calculations, we shall take the
regular solution of the Liouville equation as the canonical
transformation excluding the interesting singular part [6] which
can be treated by regular methods [7], too. But in any case locality
plays an important role [5,8].

We are going to discuss the non-critical bosonic string theory as a
particularly simple and interesting example. It induces the non-linear
Liouville dynamics as an anomaly [9] and describes for space-time
dimension $d=0$ pure 2-dimensional gravity, for $d=25$ the critical string,
and in between matter coupled to gravity.

We shall follow in this talk mainly references [5,10].
 First we look at the classical structure of
Liouville theory and search for a
suitable free field which allows us to
construct a Liouville vertex operator. But
canonical quantization requires  deformations in order to satisfy the
conformal and local properties which characterize the
classical Liouville theory, otherwise unwanted anomalies would appear.
These deformations indicate an internal $SL(2)_q$
quantum group structure.

We calculate the algebraic properties of the quantum Liouville theory and
discuss correlation functions
of non-critical strings for the particularly interesting space-time
dimensions $d=1$ and $d=25$, where as a function of d the Liouville
field behaves
either as an euclidean `coordinate' or as a minkowskian `time' respectively.
Finally, some conclusions are drawn.

\vspace{2.0cm}

\noindent
{\Large \bf 2.\ \ Canonical quantization of Liouville theory}

\vspace{1.0cm}

\noindent
{\large \bf 2.1 The classical structure}

\vspace{1.0cm}

We choose  canonical quantization of Liouville theory as
an alternative to handle the still unsolved path integration over the
exponential Liouville action ($D_t \varphi$ is the translation invariant
measure)

\begin{equation}
<\prod_{k}O_{k} (\varphi)> \equiv \int D_{t}\varphi
\prod_{k}O_{k}(\varphi) e^{\frac{d-25}{48\pi}
\int ((\partial\varphi)^{2}+\mu^{2}e^{\varphi})}
\end{equation}

\noindent
The calculations will be carried out in the conformal gauge

\begin{equation}
g_{ab} = e^{\varphi(\tau,\sigma)} \delta_{ab}
\end{equation}
\noindent
for the simple genus zero case and periodic bounderies

\begin{equation}
\varphi(\tau,\sigma) = \varphi(\tau,\sigma+2\pi)
\end{equation}

In order that the quantization feels the non-linear Liouville dynamics,
we will  understand the solution of the Liouville
equation $(\xi_{\pm} = \tau \pm \sigma)$

\begin{equation}
e^{\lambda\varphi(\tau,\sigma)} = \left( \frac{1}{\sqrt{A'(\xi_{+})}}
\frac{B(\xi_{-})}{\sqrt{B'(\xi_{-})}} -
\frac{A(\xi_{+})}{\sqrt{A'(\xi_{+})}} \frac{1}
{\sqrt{B'(\xi_{-})}} \right)^{-2\lambda}
\end{equation}

\noindent
as a canonical transformation between the Liouville field
$\varphi(\tau,\sigma)$ and a canonical free field $\psi(\tau,\sigma)$

\begin{equation}
\psi(\tau,\sigma) = \psi^{+}(\xi_{+}) + \psi^{-}(\xi_{-})
\end{equation}

Indeed, the chiral functions

\begin{equation}
\psi_{-\frac{1}{2}}(\xi_{+}) = (A'(\xi_{+}))^{-\frac{1}{2}}, \qquad
\psi_{+\frac{1}{2}}(\xi_{+}) = A(\xi_{+})(A'(\xi{+}))^{-\frac{1}{2}}
\end{equation}

\noindent
respectively the anti-chiral ones

\begin{equation}
\chi_{-\frac{1}{2}}(\xi_{-}) = (B'(\xi_{-}))^{-\frac{1}{2}}, \qquad
\chi_{+\frac{1}{2}}(\xi_{+}) = B(\xi_{-})(B'(\xi{-}))^{-\frac{1}{2}}
\end{equation}
can easily be expressed by free fields. We remind the reader [3]
that $\psi_{\pm\frac{1}{2}}(\chi_{\pm\frac{1}{2}})$ behave
under $SL(2,R)$ transformations

\begin{equation}
A(\xi_{+}), \quad B(\xi_{-})\quad \rightarrow \quad T[A] = \frac{aA+b}{cA+d},
\quad T[B]
\end{equation}

\noindent
as fundamental representations $(-\frac{1}{2},0)$ $((0,-\frac{1}{2}))$,
which satisfy the Fuchsian differential equation

\begin{equation}
\left( \frac{d^{2}}{d\xi^{2}_{+}} + \frac{1}{2}T(\xi_{+}) \right)
\psi_{\pm\frac{1}{2}}(\xi_{+}) = 0
\end{equation}

\noindent
(and correspondingly for $\chi_{\pm\frac{1}{2}}(\xi_{-})$).
 $\varphi(\tau,\sigma)$ remains invariant.
$T(\xi_{+})\ (T(\xi_{-}))$ is the chiral ( anti-chiral ) component
of the energy-momentum-tensor given by the Schwarzian
derivative of $A\ (B)$.

An exponential free field representation of
these chiral fields

\begin{equation}
\psi_{\pm\frac{1}{2}} \sim e^{-\frac{1}{2}\psi^{+}}, \qquad
\chi_{\pm\frac{1}{2}} \sim e^{-\frac{1}{2}\psi^{-}}
\end{equation}

\noindent
could be the solution of the problem since
$T(\xi_{\pm})$ becomes the  improved free field
energy-momen\-tum tensor

\begin{equation}
T(\xi_{\pm}) = -\frac{1}{2}(\partial\psi^{\pm})^{2} +
\partial^{2}\psi^{\pm}
\end{equation}

\noindent
However this is not the canonical transformation we are looking for.
In this case we would have to
interpret  (4) as a
$SL(2,R)$ decomposition of a tensor power of the fundamental representation
$(-\frac{1}{2},-\frac{1}{2})$ by infinite-dimensional representation theory.
But this is not an easy and  straightforward task [11].

We can avoid this difficulty if we assume the exponential form (10) for
$\psi_{-\frac{1}{2}}(\chi_{-\frac{1}{2}})$ only and
find $\psi_{+\frac{1}{2}}(\chi_{+\frac{1}{2}})$, the
second solutions of the
Fuchsian differential equations, by integration
using the monodromy properties for
closed strings

\begin{eqnarray}
A(\tau +  \sigma + 2\pi)   & = e ^{\gamma p/2} A(\tau + \sigma) \nonumber\\
                           &                                   \\
B(\tau -  (\sigma + 2\pi)) & = e ^{\gamma p/2} B(\tau - \sigma) \nonumber
\end{eqnarray}

\noindent
We obtain

\begin{equation}
A(\tau + \sigma) = \frac{1}{2} \sinh^{-1} \left(\frac{1}{4}\gamma p \right)
\int^{2\pi}_{0}d
\sigma'e^{\frac{1}{4}\gamma
p\epsilon(\sigma-\sigma')}e^{\psi^{+}(\tau+\sigma')}
\end{equation}
\noindent
(and correspondingly for $B (\xi_{-})$). The Liouville solution
(4) then becomes for $\lambda=-1/2$ the well-known
integrated canonical B\"{a}cklund transformation [4]

\begin{eqnarray}
e^{-\frac{1}{2}\varphi(\tau,\sigma)} = & e^{-\frac{1}{2}(\psi^{+}(\tau+\sigma)
-\psi^{-}(\tau-\sigma))} \sinh^{-1}(\frac{1}{4}\gamma p) \int^{2\pi}_{0}d
\sigma'e^{\frac{1}{4}\gamma p\varepsilon(\sigma-\sigma')} \nonumber\\
                               &                         \\
                               & \cdot  e^{\frac{1}{2}(\psi^{+}(\tau+\sigma')
-\psi^{-}(\tau-\sigma'))} \cosh(\frac{1}{2}\psi(\tau,\sigma')) \nonumber
\end{eqnarray}

\noindent
Although this compact expression allows to some extent exact
quantum-mechanical
calculations [4],it is equally useless for defining a Liouville vertex
operator $e^{\lambda\varphi}$.

Motivated by ref.[12], we found a more suitable canonical
transformation of the Liouville field
in terms of the free field

\begin{equation}
\psi(\tau,\sigma) = \ln \left( A'(\xi_{+})(-\frac{1}{B(\xi_{-})})' \right)
\end{equation}

\noindent
It exponentiates  $( B(\xi_{-})$ is here replaced by $-1/B(\xi_{-})$
in a formula like (13) !)

\begin{equation}
\psi_{-\frac{1}{2}}(\xi_{+}) = e^{-\frac{1}{2}\psi^{+}(\xi_{+})}, \qquad
\chi_{+\frac{1}{2}} = e^{-\frac{1}{2}\psi^{-}(\xi_{-})}
\end{equation}

\noindent
and allows to factorize the Liouville solution (4)  into the
form

\begin{equation}
e^{\lambda\varphi(\tau,\sigma)} = e^{\lambda\psi(\tau,\sigma)}(1-A(\xi_{+})
/B(\xi_{-}))^{-2\lambda}
\end{equation}

The free field factor $e^{\lambda\psi}$ now carries the total
conformal weight of
$e^{\lambda\varphi}$, whereas the bracket of (17) has conformal weight zero.
The
infinite-dimensional representation theory of $SL(2,R)$ does not
operate here, and the canonical transformation

\begin{equation}
e^{\lambda\varphi(\tau,\sigma)} = e^{\lambda\psi(\tau,\sigma)}
\,(1 + \mu^{2} X(\tau,\sigma))^{-2\lambda}
\end{equation}

\noindent
can, at least formally, be expanded as

\begin{equation}
e^{\lambda\varphi(\tau,\sigma)} = \sum^\infty_{n=0} \frac{(-\mu^2)^n}{n!}
\frac{\Gamma(2\lambda + n)}{\Gamma(2\lambda)}
Z^{(\lambda,n)}(\tau,\sigma)
\end{equation}

\noindent
where we used the notations

\begin{equation}
Z^{(\lambda,n)}(\tau,\sigma) =
e^{\lambda\psi(\tau,\sigma)} (X(\tau,\sigma))^n
\end{equation}

\noindent
and

\begin{equation}
X(\tau,\sigma) = \frac{1}{4} \sinh^{-2}({\frac{1}{4}} \gamma p)
\int^{2\pi}_0 d\sigma' d\sigma'' e^{\frac{1}{4}\gamma p
(\varepsilon(\sigma - \sigma') - \varepsilon(\sigma - \sigma''))}
e^{\psi^+ (\tau + \sigma ')} e^{\psi^- (\tau - \sigma ')}
\end{equation}

\noindent
In the following we shall construct a Liouville vertex operator by
means of the expansion (19).

\vspace{1.0cm}

\noindent
{\large \bf 2.2 The quantum structure}

\vspace{1.0cm}

This section mainly follows reference [5]. Once the Liouville fields
are expressed in terms of  free fields

\begin{equation}
\psi^{\pm} (\xi_{\pm}) = \frac{1}{2} \gamma q +
\frac{\gamma}{4\pi} p \xi_{\pm} + \frac{i\gamma}{\sqrt{4\pi}}
\sum_{n \neq 0} \frac{1}{n} a_n^{\pm} e^{-in \xi_{\pm}}
\end{equation}

\noindent
we define the
corresponding Liouville operators by
canonically quantizing these free fields, as usual, by means of the
commutation relations

\begin{equation}
[q,p] = i  ,\qquad [a^{\pm}_n, a^{\pm}_m] =n\delta_{n+m,o}\qquad , ...
\end{equation}

\noindent
and take into account the normal-ordering prescription. But  we have to pay
attention that the
conformal covariance
of the classical Liouville theory

\begin{equation}
\left[L_n , e^{\lambda \varphi (\sigma)} \right] =
-e^{in\sigma} (i \partial_{\sigma} + n \Delta(\lambda))
e^{\lambda\varphi(\sigma)}
\end{equation}

\noindent
and locality

\begin{equation}
\left[e^{\lambda\varphi(\sigma)} , e^{\delta\varphi (\sigma')} \right] = 0
\end{equation}

\noindent
remain valid quantum-mechanically, too.
Here $L_{n}$ is the Liouville-Virasoro
operator and $\Delta(\lambda)$ the conformal weight of $e^{\lambda\varphi}$.

Naive calculations would provide anomalous contributions to (24, 25).
To get rid of them we have to accept
quantum deformations [3-5].
Let us  briefly illustrate the situation by writing, without proof,
the most simple explicit example. It is given by the chiral fields
$\psi_{\pm\frac{1}{2}} (\chi_{\pm\frac{1}{2}})$
themselves which quantum-mechanically become

\begin{eqnarray}
\psi_{-\frac{1}{2}}(\sigma) & = & : e^{-\frac{1}{2}\eta\psi^{+}(\sigma)}:
\nonumber \\
\psi_{+\frac{1}{2}}(\sigma) & = & \frac{1}{2} : \sinh^{-1} \left( \frac{1}{4}
\gamma\eta p +
2 i \pi\hbar \eta^{2} \right) \psi_{-\frac{1}{2}}(\sigma) \\
  &   & \cdot \int_0^{2\pi} d \sigma' e^{\frac{1}{4}\gamma\eta p\varepsilon
(\sigma-\sigma')} \left[ 4\sin^2 \left( \frac{\sigma-\sigma'}{2} \right)
\right] ^{2\hbar\eta^2}
\psi^{-2}_{-\frac{1}{2}}(\sigma'):\;\; . \nonumber
\end{eqnarray}
\noindent
We see that the Liouville coupling $\gamma$
is renormalized

\begin{eqnarray}
\gamma \rightarrow \gamma \eta \;\; ,
\end{eqnarray}

\noindent
find the $\sinh$ factors quantum-mechanically shifted in the
momentum zero-mode with
respect to the classical expression (13), and observe short-distance
factors that were eliminated by the normal-ordering prescription
reintroduced here as a conformal correction.

In case that the renormalization parameter $\eta$ obeys the quadratic
equation

\begin{eqnarray}
2\hbar\eta^2 - \eta + 1 = 0 , \quad \hbar = \gamma^2 / (16 \pi)
= 3 / (25-d)
\end{eqnarray}

\noindent
we could show that
these quantum deformed chiral operators (26) satisfy
the quantum Fuchsian differential equation [3]

\begin{eqnarray}
\left( \frac{d^2}{d\sigma^2} - \frac{1}{4} \hbar\eta^2 \right)
\psi_{\pm \frac{1}{2}} =2 \hbar \eta^2 \Bigg[\sum^{\infty}_{n=1}
e^{in\sigma} L_{-n}\psi_{\pm \frac{1}{2}}
L + \frac{1}{2} L_{0} \psi_{\pm\frac{1}{2}}
\nonumber\\
+ \frac{1}{2} \psi _{\pm\frac{1}{2}} L_{0} + \psi_{\pm\frac{1}{2}}
\sum^{\infty}_{n=1} e^{-in\sigma} L_n \Bigg]
\end{eqnarray}
\noindent
(the same holds for $\chi_{\pm \frac{1}{2}}$), and that all
these  operator fields have the same conformal weight

\begin{equation}
\Delta \left( - \frac{1}{2} \right) = -\frac{1}{2} \left (\eta +
\hbar\eta^2 \right)
\hbox{\kern1pt\lower1ex\hbox{$\buildrel\longrightarrow\over
{\scriptscriptstyle\hbar\to 0}$}} - \frac{1}{2}
\end{equation}

\noindent
We should also mention here that
 $\psi_{+\frac{1}{2}}$ and $\chi_{-\frac{1}{2}}$ are defined in a
reparametrization invariant way by  the same condition (28), which means that
the
expression under the integral of (26) has conformal weight one

\begin{equation}
\Delta(1) = \eta - 2 \hbar \eta^2 = 1
\end{equation}

\noindent
This quadratic equation determines $\eta$ as a function of
the space-time dimension $d$

\begin{equation}
g = 2\hbar\eta^2_{\mp} = \eta_{\mp} -1 = \frac{1}{12} \left(13-d \mp
\sqrt{(1-d)(25-d)}\right)
\end{equation}

\noindent
The parameter $g$ controls any quantum correction in the string-induced
quantum Liouville theory. It also determines the KPZ critical
exponents and the anomalous Kac-Moody central charge of SL(2,R).
But $g$ underlies here several serious limitations, for instance,
reality of the conformal weight of physical operators
restricts the space-time dimension to

\begin{equation}
d\le 1 \qquad or \qquad d\ge 25
\end{equation}

\noindent
excluding the physical interesting region $1<d<25$ where g becomes complex.

Now we are ready to discuss
the Liouville vertex operator
$e^{\lambda\varphi (\sigma)}$.
 Although possessing the operators
$\psi_{\pm \frac{1}{2}} (\sigma)$
$\left(\chi_{\pm \frac{1}{2}} (\sigma)\right)$, we cannot simply calculate the
vertex operator from them as in the classical case.
The composite operator
 $e^{\lambda\varphi(\sigma)}$ has to be constructed separately by
means of the whole machinery of
quantization rules which we have disscussed before.
We obtain, finally, for the Liouville vertex operator a remarkable result
( from here on  $\eta$ is absorbed in $\psi$ ! )

\begin{eqnarray}
e^{\lambda\varphi (\sigma)}& = &: e^{\lambda\psi (\sigma)} :
\sum_{n=0}^{\infty} \frac{(-\mu^2)^n}{[n!]_q}
\frac{\Gamma_q (2\lambda +n)}{\Gamma_q (2\lambda)}\nonumber\\
& &\prod^n _{j=1} F \left(\lambda+j-1, \frac{1}{4}\gamma \eta p
- 2 \pi i
(\lambda-n)\hbar \eta^2 \right) \bigg[:S(\sigma):\bigg]^n
\end{eqnarray}

We recognize the special normal-ordering as a consequence of the
requirement that the
conformal covariance condition (24) should be anomaly-free.
It hides the reintroduced short-distance singularities

\begin{eqnarray}
\left[ 4 \sin^2 \frac{\sigma'_i -\sigma'_j}{2} \right]^{-g}
\end{eqnarray}

\noindent
which become integrable by restricting the Liouville coupling $\gamma$
respectively the parameter $g$. The n-th power of the normal-ordered
`screening charge' operator in (34)

\begin{equation}
S(\sigma) = \frac{1}{4}\int^{2\pi}_0 d\sigma' d\sigma''
e^{\frac{1}{4} \gamma \eta p \left(\epsilon(\sigma-\sigma') -
\epsilon(\sigma-\sigma'')\right)} e^{\psi^+ (\sigma')}
e^{\psi^- (\sigma'')}
\end{equation}

\noindent
provides no other singularities, in particular there do not arise
singularities in the variable $\sigma$ which is an argument of
the sign-function $\epsilon(\sigma-\sigma')$ only.

Most interestingly, the additional requirement of locality (25)
does not only shift the zero-mode arguments of the
$\sinh$ factors

\begin{equation}
F(\lambda ,P) = \sinh^{-1}(P + i\pi \lambda g) \sinh^{-1} (P-i\pi \lambda g)
\end{equation}

\noindent
as expected. Here,
in addition,
numbers become q-numbers

\begin{eqnarray}
n! &\rightarrow& [n!]_q = \prod^n_{j=1} \frac{\sin(2\hbar\eta^2 j\pi)}
{\sin(2\hbar\eta^2 \pi)}\nonumber \\
\Gamma(x) &\rightarrow& \Gamma_q (x)
\end{eqnarray}

\noindent
if we turn the classical expansion (19) into its operator form (34).
This indicates a quantum group structure as a consequence of both the
conformal and
local properties of the quantum Liouville theory.

The locality condition (25) which operates iteratively for each
order n of the expansion (34)

\begin{eqnarray}
\frac{\Gamma(2\lambda+n)}{\Gamma(2\lambda)}
\bigg[\: :Z^{(\lambda,n)}_q(\sigma) : \quad,\quad
: e^{\delta\psi(\rho)} : \: \bigg]
- \big( (\lambda,\sigma) \leftrightarrow (\delta,\rho) \big)
\nonumber \\
=\sum^{n-1}_{k=1} {n \choose k} {\frac{\Gamma(2\lambda+n-k)}
{\Gamma(2\lambda)}} {\frac{\Gamma(2\delta + k)}{\Gamma(2\delta)}}
\bigg[\: : Z^{(\lambda,n-k)}_q :\quad ,\quad : Z^{(\delta,k)}_q : \: \bigg]
\end{eqnarray}

\noindent
determines the deformations (37,38) altogether. $Z_q$ is the deformed $Z$
of eq (20).

It also seems to be worth mentioning that
for the special value $\lambda = \frac{1}{2}$ the q-deformed
expansion (34) can be resummed again (the q-numbers disappear!), and
we obtain

\begin{equation}
e^{\frac{1}{2}\varphi (\sigma)}= : e^{\frac{1}{2}\psi (\sigma)} :
\left(1 - : F(\frac{1}{2}, \frac{1}{4}\gamma\eta p-\pi i\hbar\eta^2)
S(\sigma):\right)^{-1}
\end{equation}

\noindent
The short-distance singularities of the expansion (34) which are contained in
$[:S(\sigma):]^n$ surprisingly disappear
in the analytic result (40). This may indicate
that expansions like (34) have only a very formal nature.

Finally, we make reference to the conformal weight
of the operator (34)

\begin{equation}
\Delta(\lambda) = \lambda\eta (1-2\hbar\eta\lambda) = \bar{\lambda}\eta
(1-2\hbar\eta\bar{\lambda})
\end{equation}

\noindent
which defines a `background charge' $2\beta_0$

\begin{equation}
\bar{\lambda} = 2\beta_0 - \lambda, \qquad 2\beta_0 = \frac{1}{2\hbar\eta},
\qquad \Delta (2\beta_0) = 0
\end{equation}

\noindent
The Liouville vertex operator so has the expected short distance
behaviour $(z = e^{i\sigma})$

\begin{equation}
e^{\lambda\varphi (z)} e^{\bar{\lambda}\varphi (z')} \sim
\mid z-z' \mid^{-2\Delta(\lambda)}
\end{equation}

\noindent
It satisfies, furthermore, the operator Liouville equation and represents,
indeed, a canonical transformation between $\varphi$ and $\psi$,
classically as well as quantum-mechanically [5].

\vspace{1.0cm}

\noindent
{\large \bf 2.3 The exchange algebra}

\vspace{1.0cm}

It is obvious in this canonical quantization  that the exchange
algebra of the chiral (anti-chiral) fields
$\psi_{\pm\frac{1}{2}} (\chi_{\pm\frac{1}{2}})$

\begin{equation}
\psi_{i} (\sigma) \psi_{k} (\rho) = S^{r s}_{i k}
\psi_{r} (\rho) \psi_s (\sigma) \quad, \ldots \nonumber
\end{equation}

\noindent
can be calculated by direct operator algebra. We obtain for the
exchange matrix S a result
as in ref. [3], which shows that both the canonical quantization
and the quantization by the quantum group are equivalent.

The matrix S unfortunately depends on the momentum
zero-mode and local phases

\begin{eqnarray}
S^{nn}_{nn} & = & e^{-i\pi\hbar\eta^2 \varepsilon(\sigma-\rho)} \quad ,
\qquad \qquad n=1,2 \nonumber \\[0.5ex]
S^{12}_{12}     & = & ie^{-\frac{1}{4}\gamma\eta p + i \pi\hbar\eta^2}
\frac {\sin(\pi\hbar\eta^2)}{\sinh(\frac{1}{4} \gamma\eta p)}
e^{-i\pi\hbar\eta^2 \varepsilon(\sigma-\rho)} \\[0.5ex]
S^{21}_{12}    & = & \left(1 - S^{12}_{12} \right)
\frac{\sinh(\frac{1}{4} \gamma \eta p + 2 i\pi\hbar\eta^2)}
{\sinh(\frac{1}{4} \gamma \eta p)}
e^{-i\pi\hbar\eta^2 \varepsilon(\sigma- \rho)} \quad, \ldots \nonumber
\end{eqnarray}

\noindent
But in a different basis [13]

\begin{equation}
\xi_k = \sum_{i= \pm \frac{1}{2}} u^i_k \psi_i  \quad, \quad
\zeta_k =\sum_{i=\pm\frac{1}{2}} v^i_k \chi_i \quad, \quad
k = \pm \frac{1}{2}
\end{equation}

\noindent
the new exchange matrix S will become a $SL(2)_q$ R-matrix, showing
more directly the
 internal
quantum group structure of Liouville theory.

The local phases in (45) arise because of the symmetric distribution of the
zero modes among
the left and right moving fields in $\psi, \chi$ (compare eq.
(21))
and would
disappear in the exchange algebra of the fields
$\pi_{\pm} = : \psi_{\pm\frac{1}{2}} \chi_{\mp\frac{1}{2}} :$ .

\vspace{2.0cm}

\noindent
{\Large \bf 3.\ \ Correlation functions for non-critical strings}

\vspace{1.0cm}

The calculation of correlation functions is the most difficult and
likewise uncertain problem in the quantum Liouville theory. There does
neither exist a M\"{o}bius invariant vacuum nor `screening charge'
conservation, tools which often determine the concrete calculations and so
the results. It is also not desirable to put to zero the
cosmological term of the Liouville action $(\mu^2 =0!)$ and  replace the
Liouville theory by a free field theory with a background charge [14,15].

Correlation functions can be calculated from
reparametrization
invariant operators. For non-critical strings scattering of physical states
is  described by integrated vertex operators

\begin{equation}
V(\lambda_{k},p_{k}) = \int d^{2}z e^{\lambda_{k}\varphi(z)} e^{ip_{k}
\cdot x(z)}
\end{equation}

\noindent
Such operators are defined in a reparametrization invariant way if the total
conformal weight of the Liouville vertex $e^{\lambda_{k}\varphi}$ and the
string vertex $e^{ip_{k} \cdot x}$ is equal to one

\begin{equation}
\Delta(e^{\lambda_{k}\varphi}) + \Delta(e^{ip_{k} \cdot x}) = 1
\end{equation}

\noindent
This relates the parameters $\lambda_{k}, p_{k}$ to the spacetime
dimension  $d$

\begin{equation}
\lambda_{k} = \left(\sqrt{25-d} - \sqrt{1-d+12\vec{p}_{k}^{2}}\right)
/ \left(\sqrt{25-d} - \sqrt{1-d}\right)
\end{equation}

As a simple example, we discuss a 4-point correlation function
for special `charges'

\begin{equation}
V_{4} = < P' \mid \Pi^{4}_{k=1}  V(\lambda_{k},p_{k}) \mid P>
\end{equation}

\begin{eqnarray}
\lambda_{i} = \lambda,\;  i=1,2,3;\; \lambda_{4} = \bar{\lambda} \quad with
\quad
3\lambda + \bar{\lambda} = 2\lambda + 2 \beta_{0}
\end{eqnarray}

\noindent
The ground state is chosen to be  M\"{o}bius non-invariant
with a non-vanishing momentum

\begin{equation}
\mid P> = e^{P\gamma\eta q} \mid 0>
\end{equation}
\noindent
We will use both solutions $\eta_{\pm}$ of the quadratic equation
(28) with the properties

\begin{equation}
\eta_{+} + \eta_{-} = \eta_{+}  \eta_{-} = \frac{1}{2\hbar} \ .
\end{equation}

For a while let us assume `screening charge conservation'.
Then the Liouville vertex operator  provides its own counting of
`screening char\-ges'
with the neutrality condition

\begin{equation}
2\lambda + 2\beta_{0} +  \sum_{k=1}^{4}n_{k} = 0
\end{equation}

\noindent
where the $n_{k}$  is the corresponding expansion
parameter in (34).
This neutrality condition has a non-trivial solution only if

\begin{equation}
2\lambda + 2\beta_0 = (1-m)\eta_+ + (1-n)\eta_-
\end{equation}

\noindent
is a negative integer.
There are two interesting solutions. For d= 25 the eq (32) yields
$\eta_{\pm} = 0$  and we obtain

\begin{equation}
\gamma \eta = 2i \sqrt{2\pi} , \qquad
\lambda_k^2 = 1 - \frac{1}{2} \vec{p}^2_{k}
\end{equation}

\noindent
Now the neutrality condition (55) does not allow
`screening charges' \linebreak $(n = 0$ in (34)!) and
the Liouville field $\varphi(\tau,\sigma)$ becomes the free
field $\psi(\tau,\sigma)$ with a central charge c=1 and a pure imaginary
coupling $\gamma\eta$. Since the kinetic term of the Liouville theory then
gets a negative sign, one could interprete the
Liouville field for d=25 as a time coordinate of a 26-dimensional Minkowski
space-time [16].
 For the 4-point-function we obtain then the expected critical string
result, a Virasoro-Shapiro amplitude in 26-dimensional Minkowski
space-time ($\lambda_{k}, p_{k}$ form here a new vector $p_{k}$ !)

\begin{equation}
V_4 = \int dz \mid z \mid^{\frac{1}{2} p_1 \cdot p_2} \mid 1-z
\mid^{\frac{1}{2}
p_2 \cdot p_3}
\end{equation}

The second interesting case is $d=1$, with

\begin{equation}
\eta_{\pm} = 2 , \qquad \gamma\eta = 2 \sqrt{2 \pi} , \qquad
\lambda = 1 - \frac{1}{\sqrt{2}} p
\end{equation}

\noindent
Now, the Liouville field acts like a `coordinate', and the `screening
charges' contribute to the integrand of (57) a complicated factor $K(z)$
which is partially due to the conformal corrections discussed in the
chapter before.
The integrals of these correlation functions
are in general not of the Dotsenko-Fateev type [17,10].

With such calculations we are able, for instance, to rederive results
of refs [18,19]. But we started this calculation with wrong assumptions.
Without `screening charge' conservation each term of (34) should
contribute and we should be able to sum up the whole series in order
to get analytical results.

For 3-point functions, at least, one can obtain order by order closed
results of the Dotsenko-Fateev type [20].

\vspace{2.0cm}

\noindent
{\Large \bf 4.\ \ Conclusion}

\vspace{1.0cm}

We have given an operator representation of the string induced quantum
Liouville theory. It is based on a canonical free field quantization which
preserves the non-linear Liouville dynamics and
takes into account locality and the conformal properties of Liouville
theory in a self-consistent manner.

 From the algebraic point of view, this operator formulation
 fits rather well into the general
scheme of a 2-dimensional conformal quantum field
theory, in particular, it allows to construct a Liouville vertex operator,
and it shows beyond it an interesting
internal $SL(2)_q$ quantum group structure. At least formally the algebraic
calculations are
not sensitive to the restrictions of space-time (33).

But the calculation of correlation functions
is less convincing, except that the critical string is reproduced by
interpreting the Liouville field as a time coordinate. For these
calculations we have used rather
formal expansions. It might be that in this way extra difficulties
arise,as the experience with the example (40) seems to tell us.

Finally, there also
remains  the challenge to understand the structure of the Hilbert space
of Liouville theory. The question arises if the internal quantum
group may be important in this respect.

\vspace{1.0cm}

\noindent
{\bf Acknowledgements}: I would like to thank the organizers of the Johns
Hopkins Workshop for the invitation and for their hospitality.
\vspace{2.0cm}

\noindent
{\Large \bf References}
\vspace{0.5cm}
\begin{description}
\item[[ 1]] J. Liouville, J. Math. Pure Appl. 18(1853)71.
\item[[ 2]] For a recent review see I.R. Klebanow, String Theory in Two
Dimensions, in String Theory and Quantum Gravity '91, Proceedings of
the Trieste Spring School and Workshop, Ed. J.Harvey et al, World
Scientific, Singapore 1992.

\item[[ 3]] J.-L. Gervais and A. Neveu, Nucl. Phys. B199(1982)59;
B209(1982)125;\\
B238(1984)396.
\item[[ 4]] E. Braaten, T. Curtright, and C. Thorn, Phys. Rev. Lett. 51(1983)
19; Ann. Phys. (NY)147(1984)365.
\item[[ 5]] H.J. Otto and G. Weigt, Phys.Lett.B159(1985)341; Z.Phys C31(
1986)219;\\
G. Weigt, Zeuthen Preprint PHE-90-15.
\item[[ 6]] G.P. Dzhordzhadze, A.K. Pogrebkov and M.C. Polivanov,
Teor.Math.Fiz.40(1979)221.
\item[[ 7]] L. Johansson and R. Marnelius, Nucl.Phys.B254(1985)201;\\
R. Marnelius, Nucl.Phys.B261(1985)319.
\item[[ 8]] J. Balog and L.Palla, Phys.lett.B274(1992)323.
\item[[ 9]] A.M. Polyakov, Phys. Lett. B103(1981)207.
\item[[10]] G. Weigt, Phys.Lett.B277(1992)79.
\item[[11]] F. Smirnoff and L. Takhtajan, Leningrad Steklov Preprint
(1990).
\item[[12]] E. D'Hoker and R. Jackiw, Phys,Rev.D26(1982)3517;
Phys Lett.50(1983)1719.
\item[[13]] O. Babelon, Phys. Lett. B215 (1988) 529.
\item[[14]] F. David, Mod.Phys.Lett.A3(1988)1651.
\item[[15]] J. Distler and H.Kawai, Nucl.Phys.B321(1989)509.
\item[[16]] S.R. Das, S. Naik, and S.R. Wadia, Bombay Preprint TIFR-TH-88/58
(1988).
\item[[17]] V.S.Dotsenko and V.A.Fateev, Nucl.Phys.B240(1984)312.
\item[[18]] A.M. Polyakov, Mod. Phys. Lett A6(1991)635.
\item[[19]] M.Goulian and M. Li, Phys. Rev. Lett. 66 (1991)2051.
\item[[20]] J. Schnittger, unpublished calculations.
\end{description}

\end{document}